\documentclass[debug]{epl}
\usepackage{latexsym,amsmath} 

\title{Charge order in  Magnetite. An LDA+$U$ study}
\author{Georg K. H. Madsen\inst{1,2} \and Pavel Nov{\'a}k\inst{2}}
\institute{
  \inst{1} Dept. of Chemistry, University of Aarhus, DK-8000 \AA rhus C,  Denmark.georg@chem.au.dk \\
  \inst{2} Institute of Physics, Academy of Sciences of the Czech Republic, Cukrovarnick{\'a} 10, 162 53 Praha 6, Czech Republic
}

\begin{document}
\maketitle
\date{\today}
\begin{abstract}
The electronic structure of the monoclinic structure of Fe$_3$O$_4$ is studied using both the local density approximation (LDA) and the LDA+$U$. The LDA gives only a small charge disproportionation, thus excluding that the structural distortion should be sufficient to give a charge order. The LDA+$U$ results in a charge disproportion along the $c$-axis in good agreement with the experiment. We also show how the effective $U$ can be calculated within the augmented plane wave methods.
\end{abstract}
\section{Introduction}
Magnetite has received a lot of attention both for fundamental and technological reasons. Above the Verwey transition it is a half-metal and has the highest known $T_c$ of 860 K. The crystal structure is the inverse spinel structure with the formal chemical composition Fe$^{3+}_A$[Fe$^{2+}$Fe$^{3+}$]$_B$O$^{2-}_4$. The two octahedrally coordinated $B$ positions are symmetry equivalent and order antiferromagnetically with the tetrahedrally coordinated $A$ site in the cubic $Fd\bar{3}m$ spacegroup. When cooled to the Verwey transition temperature, which lies around 122-125~K depending on sample, the conductivity of magnetite drops abruptly by two orders of magnitude. Originally the structure below the Verwey transition transition was thought to be have iron cations at the $B$ sites order as Fe$^{2+}$ and Fe$^{3+}$ along the (001) planes. This turned out to be too simple a model and single crystal diffraction studies showed that the low temperature structure is monoclinic with a $\sqrt{2}a\times \sqrt{2}a\times 2a$ supercell and a $Cc$ space group symmetry. A recent NMR study resolved 8 tetrahedral and 16 octahedral environments thereby confirming the $Cc$ space group.\cite{fe3o4nmr} However, the diffracted supercell peaks are extremely weak and even a recent synchrotron diffraction study could only resolve three supercell peaks.\cite{fe3o4sync2} The structure has therefore only been refined in a $a/\sqrt{2}\times a/\sqrt{2}\times2a$ monoclinic subcell with an additional orthorhombic $Pmca$ pseudo symmetry constraint (see refs.~\cite{iizumi,fe3o4sync2} for a detailed description of the structure).

It is generally agreed that the charge order model with alternating Fe$^{2+}$/Fe$^{3+}$ layers below the Verwey transition is too simple but the full charge order is still not fully understood. A M\"ossbauer study fitted the spectrum with five components.\cite{fe3o4moss2} One corresponding to Fe$^{3+}_A$ and four to Fe$^{2+}_B$ and Fe$^{3+}_B$ on two non-equivalent octahedral sites.\cite{fe3o4moss2}  A resonant X-ray diffraction study reported that if a charge disproportionation takes place it should be below the sensitivity limit of 25\% of the experiment.\cite{fe3o4dafs} Recently it was noticed that the structural distortions lead to significantly different mean Fe$-$O distances for the octahedrally coordinated iron sites.\cite{fe3o4sync2} When these distances are used to calculate bond valence sums a charge modulation along [001] is found and the octahedral iron sites split into two groups with a charge disproportion of about 0.2 electron. Furthermore a [00$\frac{1}{2}$] second charge modulation is found, which leads to the doubling of the $c$-axis compared to the cubic structure.\cite{fe3o4sync2}

Theoretically the problem of charge order in Fe$_3$O$_4$ has been difficult to
attack. First of all the true structure is unknown and even the
simplified $a/\sqrt{2}\times a/\sqrt{2}\times2a$ subcell is highly complex for
a theoretical calculation. Furthermore, the local density approximation (LDA) is not generally
applicable to highly correlated transition metal oxides, due to the spurious self interaction. F.inst. if the
calculations are done in the originally proposed Verwey model without
structural distortions, an LDA calculation converges to a metallic state with
the octahedral iron sites in an Fe$^{2.5+}$ oxidation
state.\cite{anisimovfe3o4} In the LDA+$U$ method an orbital dependent field is
introduced which can be shown to give a correction for the self
interaction.\cite{ldaUfll} Consequently two studies have shown that the
LDA+$U$ method leads to a charge ordering.\cite{anisimovfe3o4,antonovfe3o4}
However, both these studies used the original Verwey model of the structure thereby omitting
both the structural distortions and the additional [00$\frac{1}{2}$] charge
modulation.\cite{fe3o4sync2} Till date only one calculation has
been done on the $a/\sqrt{2}\times a/\sqrt{2}\times2a$
structure.\cite{svanefe3o4} Both a pure LDA and three self interaction corrected (SIC)-LDA calculations
were performed and the authors reached two surprising
conclusions.\cite{svanefe3o4} First of all it was found that the scenario
where five $d$-electrons move in the SIC-LDA potential on both the $A$ and all
$B$ sites (which can be
interpreted as all the octahedral iron sites being in the Fe$^{3+}$ oxidation
state) was the most stable.\cite{svanefe3o4} Secondly the lowest energy state
shows only a small charge disproportion of 0.1 electron. This is of similar
magnitude to the charge disproportion found using pure LDA, leading to the
conclusion that the charge disproportion is structural of
origin.\cite{svanefe3o4} One possible objection to these results could be that
the minimal basis set and atomic sphere approximation used for the
potential\cite{svanefe3o4} could make the total energies unreliable. A second
objection could be that in the SIC-LDA method one must choose a set of
localized states for which the SIC is applied. The actual ground state will
thus only be found if it is among those tested. 

We have therefore performed LDA+$U$ calculations using the linearized augmented plane wave(LAPW) method. Furthermore we have tested a method for deriving $U$-values using the LAPW method. We do indeed find an electronic configuration that was not considered in the SIC-LDA study\cite{svanefe3o4}, but which is in good agreement with the charge disproportion that was proposed based on the experimental bond distances.\cite{fe3o4sync2} The paper is arranged into two sections. First we describe the LDA-$U$ method and how we derive $U$-values within the LAPW method. Secondly the calculated electronic structure is described.

\section{Fixing the $U$ within the LAPW method}
The LDA+$U$ method essentially consists of identifying a set of atomic like orbitals which are treated in a non-LDA manner.\cite{ldaU} Based on the lessons from Hubbard model studies these orbitals are treated with an orbital dependent potential with an associated on-site Coulomb and exchange interactions, $U$ and $J$.To avoid double counting in the non-spherical part of potential we use $U_{eff}=U-J$ instead of $U$\cite{dudarev} and omit the multipolar terms proportional to $J$ in the added LDA+$U$ potential.
\begin{equation}
E^{orb}(\hat{n}) =-\frac{U-J}{2}\sum_\sigma Tr(\hat{n}^\sigma \cdot \hat{n}^\sigma)
\label{eq:eorb1}
\end{equation}
where $\hat{n}$ is the orbital occupation matrix. This form is both rotationally invariant and is well suited for full potential calculations.

In LDA the electron-electron interactions have already been taken into account in a mean field way.  One must therefore identify the parts that occur twice and apply a double counting correction (DCC).
Several different versions of the DCC exist.\cite{ldaU,ldaUfll,ldaUamf,mazinU} In the present work we will use what has been called the fully localized limit (FLL).\cite{ldaUfll} The FLL-DCC has a clear interpretation and has also been shown to give results in better agreement with experiment than other DCC.\cite{bobcu2o}. The FLL-DCC is constructed so that the atomic like limit to the total energy is satisfied, which leads to the following expression\cite{ldaUfll}
\begin{equation}
E^{DCC}_{FLL}=-\bigl(\frac{U}{2}n(n-1)-\frac{J}{2}\sum_\sigma n^\sigma(n^\sigma-1)\bigr)=-\frac{U-J}{2}\sum_\sigma (2l+1)n^\sigma
\label{eq:efll}
\end{equation}
where $n_\sigma=Tr(\hat{n}^\sigma)/(2l+1)$. The orbital dependent potentials entering the Kohn-Sham equation that arise from the $E^{orb}-E^{DCC}$ correction to the total energy, Eq.~(\ref{eq:eorb1}) and Eq.~(\ref{eq:efll}) is then  
\begin{equation}
\Delta V^{U \sigma}_{FLL}=\frac{\partial(E^{orb}-E^{DCC}_{FLL})}{\partial n^\sigma}=-(U-J)(\hat{n}^\sigma-\frac{1}{2}I) 
\label{eq:vfll}
\end{equation}
which shows that the FLL version of the orbital potential will stabilize an orbital that is more than half occupied and destabilize an orbital that is less than half occupied.

\begin{figure}
\twoimages{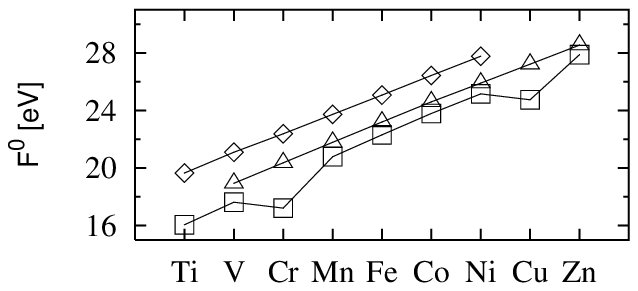}{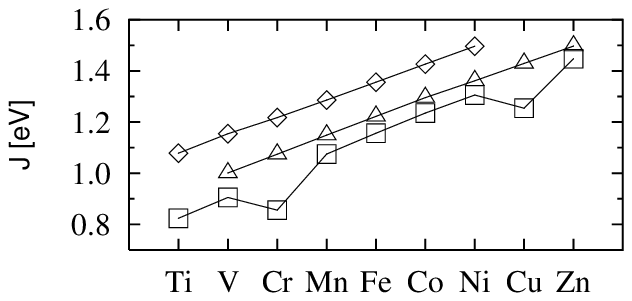}
\caption{Onsite parameters calculated from the atomic Slater integrals. $\Box$ atomic values, $\triangle$ M$^{2+}$ ions and $\diamond$ M$^{3+}$ ions. The atomic values for Cr and Cu were calculated with 5 and 10 $d$-electrons respectively.}
\label{fig:slaterint}
\end{figure}

The meaning of the $U$ parameter was discussed by Anisimov and Gunnarsson,\cite{ucalc1} who defined it as the cost in Coulomb energy by placing two electrons on the same site. In an atom the $U$ corresponds to $F^0$ of the unscreened Slater integrals.\cite{ucalc1} $F^0$ should thus both increase with increased ionicity and as the $d$-wave function is contracted across the 3$d$ transition series.  This is illustrated in Fig.~\ref{fig:slaterint} where we show the calculated atomic Slater integrals for chemically relevant 3$d$ ions.

Due to screening the effective $U$ in solids is much smaller than $F^0$ for atoms. To calculate the effective $U$ Anisimov and Gunnarsson,\cite{ucalc1} constructed a supercell and set the hopping integrals connecting the $3d$ orbital of one atom with all other orbitals to zero. The number of electrons in this non-hybridizing $d$-shell was varied and $F^0_{eff}$ was then calculated from
\begin{equation}
\begin{split}
F^0_{eff}=&\varepsilon_{3d\uparrow}((n+1)/2,n/2)-\varepsilon_{3d\uparrow}((n+1)/2,n/2-1) \\
          &-\varepsilon_F((n+1)/2,n/2)+\varepsilon_F((n+1)/2,n/2-1)
\end{split}
\label{eq:f0}
\end{equation}
where $\varepsilon_{3d\uparrow}$ is the spin-up $3d$ eigenvalue. Using the method of Anisimov and Gunnarsson, {$U$-values} have been calculated for the di- and trivalent configurations of the 3$d$ elements in La perovskites.\cite{Uperovskite} As expected the trends were the same as observed for atoms, Fig.~\ref{fig:slaterint}, but in smaller magnitude. The trends being, i), an almost linear relation between atomic number and calculated $U$ and ,ii), a constant shift between the M$^{2+}$ ions, ranging from approximately 6.5~eV (Titanium) to 8.5~eV (Copper), and the M$^{3+}$ ions, ranging from 8~eV to 10~eV.\cite{Uperovskite}

The original LDA+$U$ method\cite{ldaU} was based on the linearized muffin tin orbitals basis set, where the individual orbital and hopping terms can be identified. This is not possible within the LAPW method, so the method of Anisimov and Gunnarsson\cite{ucalc1} can not be directly applied. Instead the hybridization can be removed putting the $d$-states into the core or by performing a two-window calculation. To check this procedure we have performed calculations on the well characterized NiO.\cite{dudarev} Two calculations were performed on $2\times2\times2$ supercells each with one impurity site with the $d$-configuration forced to be as in Eq.~(\ref{eq:f0}). The $d$-character of the APWs at the impurity sites was eliminated by placing the $d$-linearization energy at a very high value. Using Eq.~(\ref{eq:f0}) we hereby got a value of $F^0_{eff}$=5.96~eV. The question is how this value should be used in an LDA+$U$ calculation? When using the spherically averaged form of $E^{orb}$, Eq.~(\ref{eq:eorb1}), $J$ simply rescales the orbital term, Eq.~(\ref{eq:efll}) and Eq.~(\ref{eq:vfll}). Furthermore Eq.~(\ref{eq:vfll}) shows that an occupied and an unoccupied orbital will be split by $U-J$. Following the interpretation of Anisimov and Gunnarsson\cite{ucalc1}, this suggest that $U-J=F^0_{eff}$. As the screening of $F^2$ and $F^4$ in solids appears to be small, $J$ can to a good approximation be calculated from the atomic values.\cite{ucalc1} Using a $J=1.36$~eV, Fig.~\ref{fig:slaterint}, we arrive at $U=7.32$~eV in good agreement with earlier values.\cite{dudarev} Furthermore an effective onsite term of $F^0_{eff}$=5.96~eV, leading to a corresponding splitting of occupied and unoccupied bands, is in excellent agreement with experiment\cite{dudarev}

We then applied the method to FeO and Fe$_2$O$_3$ and obtained $F^0_{eff}=5.73$~eV and 7.33 eV for Fe$^{2+}$ and Fe$^{3+}$ respectively. Finally calculations were performed on magnetite using the $2\times2\times2$ supercell of the cubic high temperature structure. We tried placing both Fe$^{2+}_A$ (corresponding to one calculation with 3.5 spin up and 3 spin down electrons at the impurity site and one calculation with 3.5 spin up and 2 spin down) and Fe$^{3+}_A$ at the tetrahedrally coordinated A site. Thereby values of  $F^0_{eff}=$4.79~eV and 6.33~eV were obtained. For the octahedrally coordinated B site values of $F^0_{eff}$(Fe$^{2+}_B$)=5.03~eV, $F^0_{eff}$(Fe$^{2.5+}_B$)=6.21~eV and $F^0_{eff}$(Fe$^{3+}_B$)=7.38~eV were calculated, showing good internal consistency with the FeO and Fe$_2$O$_3$ values. Using the $J$ parameters derived from the atomic values, Fig.~\ref{fig:slaterint}, we then obtain $U$(Fe$^{3+}_A$)=7.69~eV, $U$(Fe$^{2+}_B$)=6.2~eV and $U$(Fe$^{3+}_B$)=8.73~eV for magnetite. Our calculated $F^0_{eff}$ seem somewhat larger than the $U$ and $J$ values that were earlier calculated for magnetite.\cite{anisimovfe3o4} However it is not clear what oxidation state was used to calculate these values\cite{anisimovfe3o4} and furthermore it should be noted that the values for the octahedrally coordinated $B$ site are in very good agreement with what was obtained for the octahedrally coordinated Fe ions in the perovskites.\cite{Uperovskite} 

\section{Computational details}
With the $U$ values thus fixed we performed calculations on the low-symmetry structure\cite{fe3o4sync2} using the L/APW+lo method\cite{gmapwlo}, as implemented in the WIEN2k code.\cite{wien2k} Sphere sizes of 2.0~a.u. and 1.5~a.u. were used for the Fe and O sites respectively. The plane-wave (PW) cut-off was defined by min($R_{MT}$)max($k_n$)=5.7 corresponding to approximately 3700~PW. The Brillouin-zone (BZ) was sampled on a tetrahedral mesh with 100 k points (18 in the irreducible BZ). On the tetrahedrally coordinated $A$ iron sites the calculated $F^0_{eff}=U-J=6.33$~eV was used. The calculated $F^0_{eff}=6.21$~eV for Fe$^{2.5+}_B$ was used on all octahedrally coordinated $B$ iron sites so that no charge order was forced.

When discussing charge order it is necessary to define some kind of ``atomic'' charge. In the APW method these partial charges are most commonly the integrated charge within a given sphere. They can be named $q_l^t$, where $t$ refers to a given atomic sphere and $l$ to the angular quantum number. The $q_l^t$ values have the problem that they depend on the size of the atomic sphere, thus depend on a computational parameter. They are still useful for comparing atomic charges within the same calculation, if the sphere sizes are kept equal. They can however not directly be compared with the results of other calculations/experiments. One way to make the charges more transferable is to estimate the amount of charge leaking out of the sphere and use this for rescaling. To do the rescaling we have calculated the radial function of free Fe$^{2+}$ and Fe$^{3+}$ ions. By integrating the $d$-functions up to 2~a.u. (the sphere size used for iron in the present study) we get an average scaling of 1.06. In the following two valencies will be defined for the iron sites $V^t=8-q^t_2$ (the $q^t_0$ and $q^t_1$ values are less than 0.05 electrons) and $V^{t\prime}=8-1.06\times q^t_2$. As will be shown in the next section the scaling seems to be reasonable as it brings fully localized half shells up to 5 electrons.

\section{Results and discussion}
Table~\ref{tab:charge} gives the fractional charges for Fe$_3$O$_4$. It is seen that LDA gives only a very small charge disproportionation, thus excluding that the structural distortion should be sufficient to give a charge order. In contrast the LDA+$U$ calculation does give a charge disproportionation, Table~\ref{tab:charge} of approximately 0.2~electrons between the $B2$ and $B3$ (high oxidation state) and the $B1$ and $B4$ (low oxidation state) sites, thus confirming the additional [00$\frac{1}{2}$] charge modulation.\cite{fe3o4sync2} Table~\ref{tab:charge} also reports the rescaled partial charges. It is seen that the scaling brings the charge in the majority spin $d$-states of the high oxidation state iron sites close to five. This is expected and lends credibility to our scaling procedure. The scaled LDA+$U$ results in a charge disproportion of 0.2$-$0.3 electrons and the valencies are found to be smaller than what could be expected from a naive ionic picture. 
\begin{table}
\begin{tabular}{l | c c | c c | c c | c c | c c | c c |}
\hline
\hline
  & \multicolumn{2}{c |}{A1} & \multicolumn{2}{c |}{A2} & \multicolumn{2}{c |}{B1} & \multicolumn{2}{c |}{B2} & \multicolumn{2}{c |}{B3} & \multicolumn{2}{c |}{B4} \\ 
 LDA & $\uparrow$ & $\downarrow$ &  $\uparrow$ & $\downarrow$ &  $\uparrow$ & $\downarrow$ &  $\uparrow$ & $\downarrow$ &  $\uparrow$ & $\downarrow$ &  $\uparrow$ & $\downarrow$ \\
\hline
 $q^t_2$    & 1.15 & 4.41 & 1.15 & 4.41 & 4.48 & 1.10 & 4.50 & 1.03 & 4.46 & 1.10 & 4.49 & 1.06 \\
    $V^t$   &  \multicolumn{2}{c |}{2.44} & \multicolumn{2}{c |}{2.45} & \multicolumn{2}{c |}{2.42} & \multicolumn{2}{c |}{2.47} & \multicolumn{2}{c |}{2.44} & \multicolumn{2}{c |}{2.45} \\
 $m_s$      & \multicolumn{2}{c |}{-3.296} & \multicolumn{2}{c |}{-3.291} & \multicolumn{2}{c |}{3.401} & \multicolumn{2}{c |}{3.481} & \multicolumn{2}{c |}{3.379} & \multicolumn{2}{c |}{3.448}  \\
 $q^{t\prime}_2$ & 1.21 & 4.68 & 1.22 & 4.68 & 4.75 & 1.16 & 4.77 & 1.09 & 4.73 & 1.17 & 4.76 & 1.13 \\
    $V^{t\prime}$   &  \multicolumn{2}{c |}{2.10} & \multicolumn{2}{c |}{2.11} & \multicolumn{2}{c |}{2.09} & \multicolumn{2}{c |}{2.14} & \multicolumn{2}{c |}{2.10} & \multicolumn{2}{c |}{2.11} \\
\hline
 FLL & $\uparrow$ & $\downarrow$ &  $\uparrow$ & $\downarrow$ &  $\uparrow$ & $\downarrow$ &  $\uparrow$ & $\downarrow$ &  $\uparrow$ & $\downarrow$ &  $\uparrow$ & $\downarrow$ \\
\hline
 $q^t_2$         & 0.69 & 4.68 & 0.70 & 4.68 & 4.60 & 1.01 & 4.71 & 0.62 & 4.69 & 0.70 & 4.52 & 1.06 \\
    $V^t$   &  \multicolumn{2}{c |}{2.63} & \multicolumn{2}{c |}{2.63} & \multicolumn{2}{c |}{2.39} & \multicolumn{2}{c |}{2.67} & \multicolumn{2}{c |}{2.61} & \multicolumn{2}{c |}{2.42} \\
 $m_s$   & \multicolumn{2}{c |}{-4.023} & \multicolumn{2}{c |}{-4.018} & \multicolumn{2}{c |}{3.613} & \multicolumn{2}{c |}{4.114} & \multicolumn{2}{c |}{4.017} & \multicolumn{2}{c |}{3.473} \\
 $q^{t\prime}_2$ & 0.74 & 4.96 & 0.74 & 4.96 & 4.88 & 1.07 & 5.00 & 0.66 & 4.98 & 0.74 & 4.79 & 1.13 \\
    V$^{t\prime}$   &  \multicolumn{2}{c |}{2.31} & \multicolumn{2}{c |}{2.30} & \multicolumn{2}{c |}{2.05} & \multicolumn{2}{c |}{2.35} & \multicolumn{2}{c |}{2.29} & \multicolumn{2}{c |}{2.09} \\
\hline
\end{tabular}
\caption{Fractional charges, valencies and Spin ($m_s$) magnetic moments (in $\mu_B$) calculated as explained in the  computational section. The primed values have been scaled by a factor 1.06 (see the computational section for explanation). The sites are labeled as in Ref.\cite{fe3o4sync2}. The B1 and B2 sites are both an average over two symmetry-nonequivalent sites. These sites would be symmetry equivalent in an orthorhombic unit cell and as expected (the mono-clinic distortion is very small, $\beta=90.2363$\cite{fe3o4sync2}) the values calculated for these sites are very similar and therefore only one averaged value is reported.}
\label{tab:charge}
\end{table}

\begin{figure}
\onefigure[width=.95\textwidth]{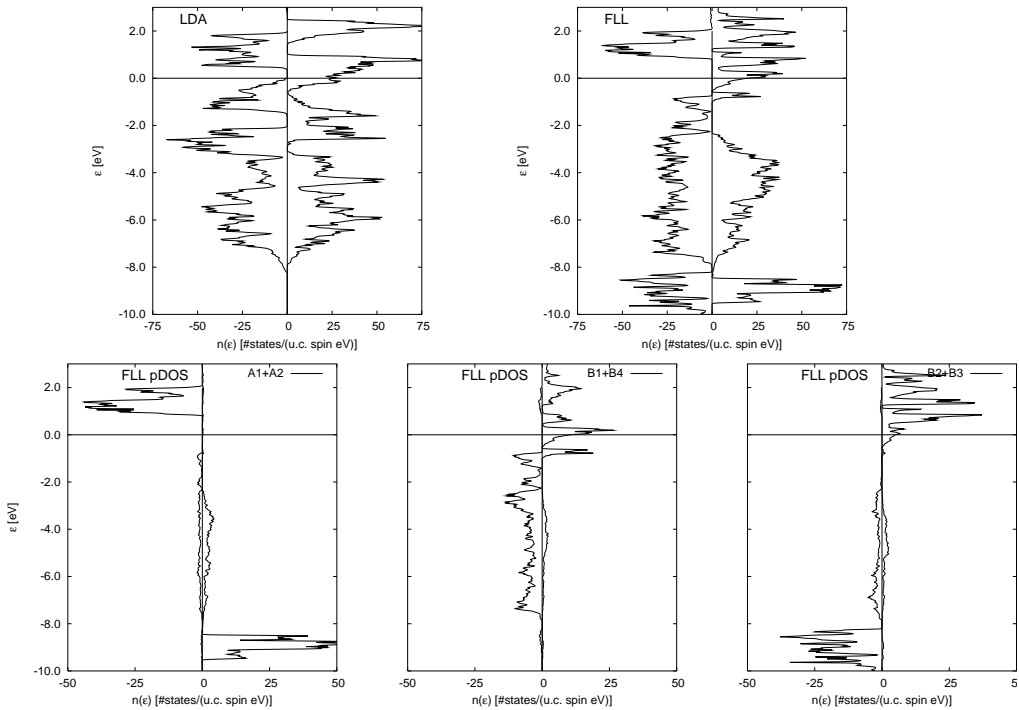}
\caption{Spin polarized density of states calculated using LDA (top left) and FLL (top right). The lower plots illustrate the partial DOS. Lower left: the $A$ (tetrahedral) sites. Lower middle: $B1$+$B4$ (octahedral with long B-O bonds) sites. Lower right: $B2$+$B3$ (octahedral with short B-O bonds) sites. }
\label{fig:fe3o4dos}
\end{figure}

Figure~\ref{fig:fe3o4dos} shows the calculated density of states (DOS) for
Fe$_3$O$_4$ in the charge ordered phase. For the LDA calculation the
oxygen band is found between approximately -4~eV and -8~eV. Above these bands one finds
the bands of Fe-3$d$ origin. Furthermore a substantial hybridization between the iron sites and
the oxygen $p$-states is seen. When the orbital field is introduced the fully
occupied minority spin bands on the $A$-iron sites are shifted
downwards below the oxygen $p$-shell. The same thing happens for the $B2$ and
$B3$ iron sites.
This was expected from Table~\ref{tab:charge}
where it is seen that the $B2$ and $B3$ have a valency similar to the $A$
sites and that the majority spin bands seem to be fully occupied. Figure~\ref{fig:fe3o4dos} also shows that the picture is very different
for the $B1$ and $B4$ sites. These sites are the ones with the long Fe$-$O
distances\cite{fe3o4sync2} and low oxidation states,
Table~\ref{tab:charge}. It is seen that there is a strong hybridization between
the majority spin Fe-3$d$ and the O-2$p$ states. Furthermore there is a
non-hybridizing minority spin states just below the Fermi level which is not
found for the $B1$ and $B4$ states and is the cause of the charge order. The
electronic structure of the low oxidation state irons sites can thus be
compared to the electronic structure of FeO,\cite{fangfeo} where the Fe-3$d$
states also show a strong hybridization with the O-2$p$ states and one sub-band
in the minority spin is occupied to accommodate the additional electron.

The experimental work\cite{fe3o4sync2} chose to renormalize their bond valence sums so that the average valency of the $B$ sites was 2.5, which is close to the average unscaled valency, Table~\ref{tab:charge}, in the present study. So despite the derived charges being method dependent (see computational section) the charges should be comparable. Consequently the valencies quoted 2.39, 2.61, 2.59 and 2.41\cite{fe3o4sync2} ($B1$, $B2$, $B3$ and $B4$ site respectively) which are in extremely good agreement with the present, Table~\ref{tab:charge}. This we believe strongly supports our
electronic structure, but then the question arrises why the earlier
computational study of the charge ordered magnetite structure\cite{svanefe3o4}
did not reach the same conclusion. 

The earlier computational work\cite{svanefe3o4} reported atomic charges and spin magnetic moments for six different $B$ sites in the monoclinic phase for four different types of calculations. One calculation (labeled SIC(1)) is performed where the simple Verwey charge order is implemented, so that the $B$ sites alternate between five $d$-electrons moving in the SIC potential and six $d$-electrons moving in a SIC potential.\cite{svanefe3o4} In our labeling a simple Verwey charge order corresponds to the atoms $B1$ and $B2$ being in a low oxidation state. These sites are actually split into two sites each in the monoclinic symmetry (see comment to Table~\ref{tab:charge}) and consequently the SIC study finds four sites (in this paper labeled $B1$, $B2$, $B3$ and $B4$\cite{svanefe3o4}) that have a low oxidation state, while two
sites ( $B3$ and $B4$ in our notation, $B5$ and $B6$ in their) have a high
oxidation state. This immediately explains why these authors found the charge
ordered state to be higher in energy than the state where all the $B$ sites
had five $d$-electrons moving in a SIC potential. The simple Verwey state
considered earlier\cite{svanefe3o4} does not correspond to the charge ordering
that could be expected from the structural distortion,\cite{fe3o4sync2} which
means that the $B3$ site is calculated to be in the high oxidation state despite it having longer distances. In other words the SIC calculation have omitted the second [00$\frac{1}{2}$] charge modulation found experimentally.\cite{fe3o4sync2}

\section{Conclusion}
By electronic structure calculations using the LDA+$U$ method we have obtained a periodic charge disproportion along the $c$-axis in the monoclinic structure of Fe$_3$O$_4$. The proposed charge order is in good agreement with the one derived from experiment and we have explained the disagreement with an earlier SIC-LDA calculation.

This work was supported by the project A1010214 of the Grant Agency of the AS CR and by the Carlsberg Foundation.


\end{document}